\documentclass[aps,prl,showpacs,twocolumn,preprintnumbers,amsmath,amssymb,longbibliography]{revtex4-2}

\usepackage{graphicx}
\usepackage{dcolumn}
\usepackage{bm}
\usepackage{color}
\usepackage{amsmath}
\usepackage{upgreek}

\begin{document}
\title{Optical computation of divergence operation for vector field }

\author{Yijie Lou, Yisheng Fang, and Zhichao Ruan}
\email{ zhichao@zju.edu.cn}

\affiliation{Interdisciplinary Center of Quantum Information, State Key Laboratory of Modern Optical Instrumentation, and Zhejiang Province Key Laboratory of Quantum Technology and Device of Physics Department, Zhejiang University, Hangzhou 310027, China}\

\begin{abstract}
Topological physics desires stable methods to measure the polarization singularities in optical vector fields. Here a periodic plasmonic metasurface is proposed to perform divergence computation of vectorial paraxial beams. We design such an optical device to compute spatial differentiation along two directions, parallel and perpendicular to the incident plane, simultaneously. The divergence operation is achieved by creating the constructive interference between two derivative results. We demonstrate that such optical computations provide a new direct pathway to elucidate specific polarization singularities of vector fields.
\end{abstract}

\maketitle

\section{Introduction}

Spatially varying polarization is one of the most fundamental properties of optical vector fields \cite{qiwen2013vectorial}. Recently, more and more studies show that the polarization singularities in vector fields play the key role to understand topological physical phenomena. Especially, in a periodic dielectric slab, a vortex as the winding of far-field polarization vectors in \textbf{k} space connects to the bound state in the continuum (BIC)  and forms ultra-high-Q resonances in open systems \cite{zhen2014topological, lee2012observation,hsu2013observation, guo2017topologically, zhang2018observation, chen2019observing}, which have been exploited for BIC laser \cite{kodigala2017lasing}. Also for electromagnetic multipoles it is important to resolve such singularities in the polarization states for identifying the closed radiation channels \cite{chen2019singularities}.

In order to analyze such vector fields, spatial polarization distributions are commonly measured by mechanically rotating polarizers and wave plates, which is time-consuming and unstable to infer the polarization singularities from the intensity variation. Over the past few years optical analog computing has attracted particular attentions with the advantage of real-time and high-throughput computations \cite{caulfield2010future}. Various optical analog computing devices were designed for spatial differentiation for scalar fields simply with linear polarization, e.g. metamaterials/metasurface \cite{Silva2014performing,Abdollah15,Chizari2016, PorsNielsenBozhevolnyi14,hwang2018plasmonic,PhysRevLett.121.173004,Zhou11137,PhysRevLett.123.013901,cordaro2019high,PhysRevApplied.11.064042,huo2020photonic}, dielectric interface or periodic slabs \cite{doskolovich2014spatial,Youssefi:16,dong2018optical,guo2018photonic,Guo:18,Wu:17,PhysRevApplied.11.054033,Zhu2019Generalized,zhu2020optical,zhou2020flat}, surface plasmonic structures \cite{ruan2015spatial,zhu2017plasmonic,Fang2017On,Fang:18}. Therefore, it can be intriguing to extend the optical computing method from scalar operation to vector one, which is capable of both directly analyzing the divergence of vector fields and revealing the polarization singularities, on a single shot.

In this article, a periodic plasmonic metasurface is proposed to perform optical analog computation of divergence operation. For linearly polarized light, our previous studies have experimentally demonstrated the spatial differentiation by two different approaches. First, based on the simplest planar plasmonic structure, the spatial differentiation along the direction parallel to the incident plane can be realized by spatial mode interference when surface plasmon polariton (SPP) is excited \cite{ ruan2015spatial,Fang2017On,zhu2017plasmonic}. Second, with the spin Hall effect of light, by analyzing polarization states orthogonal to incident light, the spatial differentiation perpendicular to the incident plane can be realized during reflection at a single optical planar interface, which generally occurs at any planar interface, regardless of material compositions or incident angles \cite{Zhu2019Generalized,zhu2020optical}. Here in order to realize optical computing of divergence operation for vectorial paraxial beams, we introduce periodic slots in the metasurface and combine these two effects together.

Such a metasurface simultaneously computes spatial differentiation along the directions parallel and perpendicular to the incident plane. We enable the divergence computation by creating the constructive interference between two derivative results. Generally, for vectorial paraxial beams, the electric fields are dominant in transverse directions. Since the divergence of total electric fields are always absent for the light propagating in air, the divergence of vectorial paraxial fields is contributed by the nonzero derivative of the longitudinal components along the beam propagation direction, which corresponds to the polarization singularities of sources or sinks.

\section{Design principle of optical divergence computation}
\begin{figure*}
\centerline{\includegraphics[width=6.4in]{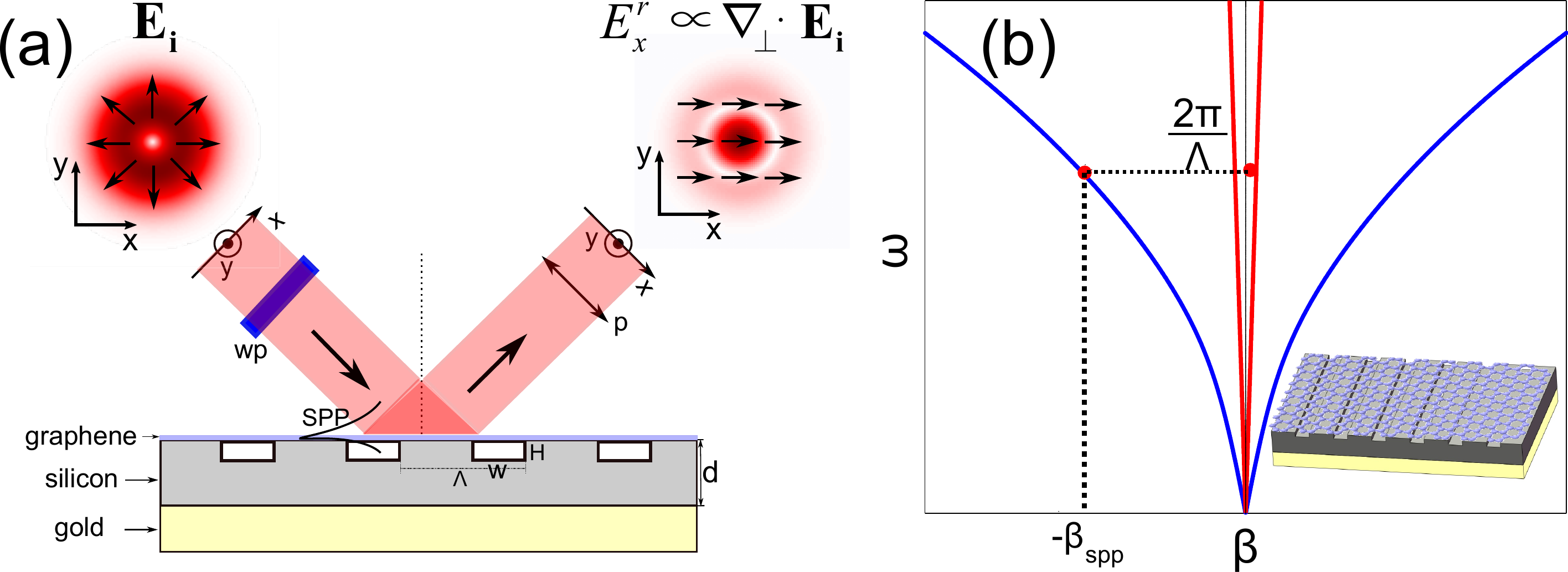}}
\caption{\label{fig:1} (a) Schematic of the periodic plasmonic metasurface for the optical computation of divergence operation. The incident angle of paraxial beam is $\theta  = {44.427^ \circ }$. The thickness of grating is $d = 5{\rm{\upmu m}}$, the period of grating $\Lambda  = 4.101{\rm{\upmu m}}$, the width of grating slot $W = 1.1323{\rm{\upmu m}}$ and the height of grating slot $H = 0.8194{\rm{\upmu m}}$.  WP: wave-plate, P: $x$ direction polarizer. The fast and slow axis of the wave plate are along the $x$ and $y$ directions, respectively.  (b) Dispersion relation of the SPPs sustained on the air-graphene-Si interface (blue line), and light line cone in Si dielectric (red line). The phase matching condition is satisfied with the help of silicon grating. The inset shows the metasurface in the three dimension.}
\end{figure*}

The periodic plasmonic metasurface is depicted in Fig.~\ref{fig:1}(a), where a silicon slot grating is fabricated on a thick gold layer substrate and then a monolayer graphene is transferred onto the grating. Here the air-graphene-silicon interface supports SPP modes which are strongly confined and propagate along the graphene layer in THz frequency region \cite{Fang2017On}. The SPP on graphene is excited through phase matching enabled by silicon grating: ${k_0}\sin \theta  - {{2\pi } \mathord{\left/{\vphantom {{2\pi } \Lambda }}\right.\kern-\nulldelimiterspace} \Lambda } =  - {\beta _{\rm{spp}}}$, where $k_0$ is the wavevector in vacuum, $\theta $ is the incident angle, ${\beta _{\rm{spp}}}$ is the wavevector of SPP. We note that here the phase matching condition ensures that only the zeroth order beam is diffracted.

We now consider that a vectorial paraxial beam illuminates on the metasurface with the incident electric fields ${{\bf{E}}_{\bf{i}}} = E_x^i\left( {x,y} \right){{\bf{e}}_{\bf{x}}} + E_y^i\left( {x,y} \right){{\bf{e}}_{\bf{y}}}$, where $x$ and $y$ are the beam coordinates perpendicular to the propagation direction, as depicted in Fig.~\ref{fig:1}. Correspondingly, the reflected beams are written as ${{\bf{E}}_{\bf{r}}} = E_x^r\left( {x,y} \right){{\bf{e}}_{\bf{x}}} + E_y^r\left( {x,y} \right){{\bf{e}}_{\bf{y}}}$. Here $E_x^{i(r)}$ and $E_y^{i(r)}$ represent $x$  and $y$ components of the vectorial paraxial fields, for the incident (reflected) beams, respectively. With the convention of $\exp \left( { - i\omega t} \right)$, each component can be decomposed to a series of plane waves: $E_\alpha ^{i\left( r \right)}\left( {x,y} \right) = \iint {\tilde E_\alpha ^{i\left( r \right)}\left( {{k_x},{k_y}} \right){e^{i{k_x}x}}{e^{i{k_y}y}}d{k_x}d{k_y}} $. Since the diffracted light of the metasurface is only in the zeroth order, each plane wave with the wavevector component $\left( {{k_x},{k_y}} \right)$  only generates the reflected plane wave with the same $\left( {{k_x},{k_y}} \right)$. Therefore, the reflected vector fields are determined by four spatial spectral transfer functions ${H_{\alpha \beta }}\left( {{k_x},{k_y}} \right)$ defined as ${H_{\alpha \beta }}\left( {{k_x},{k_y}} \right) = {{\tilde E_\alpha ^r\left( {{k_x},{k_y}} \right)} \mathord{\left/{\vphantom {{\tilde E_\alpha ^r\left( {{k_x},{k_y}} \right)} {\tilde E_\beta ^i\left( {{k_x},{k_y}} \right)}}} \right.\kern-\nulldelimiterspace} {\tilde E_\beta ^i\left( {{k_x},{k_y}} \right)}}$, where $\alpha $, $\beta $ represent $x$ or $y$ electrical components.

Next we show that the spatial spectral transfer functions ${H_{\alpha \beta }}\left( {{k_x},{k_y}} \right)$ can be engineered to realize the optical computing of divergence operation. First, when the wavevector of an incident plane wave is out of the incident plane, i.e. ${k_y} \ne 0$, the incident $y$-component field can excite the $x$-component reflected field, and vice versa, due to the transversal mode requirement of plane waves \cite{Zhu2019Generalized,zhu2020optical}. Especially when the grating slot is shallow, that is, the grating is considered as a perturbation to the air-graphene-silicon interface, under the paraxial approximation, the spatial spectral transfer function ${H_{xy}}$ has the form of
\begin{equation}
{H_{xy}} = i{A_y}{k_y}
\label{eq:1}
\end{equation}
where ${A_y}$ is a complex constant. Eq.~(\ref{eq:1}) indicates that when the incident paraxial beam is dominant along the $y$ direction ${{\bf{E}}_{\bf{i}}} = E_y^i\left( {x,y} \right){{\bf{e}}_{\bf{y}}}$, by analyzing the reflected field along the $x$ direction, the output polarized beam is $E_x^r = {A_y}\frac{{\partial E_y^i\left( {x,y} \right)}}{{\partial y}}$, and ${A_y}$ represents a scaling factor for the $y$-directional differentiation operation.

On the other hand, through phase matching, the incident electric field component in the incident plane can excite the SPP wave at the air-graphene-silicon interface. Meanwhile the excited SPP field propagating along the surface of grating leaks out and generates radiation field through grating coupling process. The total reflected field in the incident plane is contributed by the interference of direct reflected wave and leakage of SPP mode. Such a coupling process can be described by the spatial coupled-mode theory \cite{ruan2015spatial,LouPanZhuRuan16}. Especially, when the loss rate due to the SPP leaky radiation is equal to that of the material absorption, i.e. under the critical coupling condition, the total reflected field is the first-order spatial differentiation of the incident one along the $x$ direction \cite{Fang2017On}. In this situation, the corresponding spatial spectral transfer function is
\begin{equation}
{H_{xx}} = i{A_x}{k_x}
\label{eq:2}
\end{equation}
where ${A_x}$ is a complex constant determined by the loss rate due to the SPP leaky radiation.  Eq.~(\ref{eq:2}) shows that when the incident paraxial beam is dominant along the $x$ direction ${{\bf{E}}_{\bf{i}}} = E_x^i\left( {x,y} \right){{\bf{e}}_{\bf{x}}}$, by analyzing the reflected field along the $x$ direction, the output polarized beam is $E_x^r = {A_x}\frac{{\partial E_x^i\left( {x,y} \right)}}{{\partial x}}$, where ${A_x}$ represents the scaling factor for the $x$-directional differentiation operation.

Therefore, for an incident vector field  ${{\bf{E}}_{\bf{i}}} = E_x^i\left( {x,y} \right){{\bf{e}}_{\bf{x}}} + E_y^i\left( {x,y} \right){{\bf{e}}_{\bf{y}}}$, after analyzing the reflected field along the $x$ direction, the output polarized beam $E_x^r$ is the interference result between the derivatives of two incident field  components $E_x^i\left( {x,y} \right)$ and $E_y^i\left( {x,y} \right)$ along the $x$ and $y$ directions respectively. We note that generally the scaling factors ${A_y}$ and ${A_x}$ are two complex numbers with different phases. Suppose the wave plate in Fig.~\ref{fig:1}(a) has a phase delay $\Delta \phi $ in the $y$ component, the output polarized beam after analyzing the reflected fields along the $x$ direction is
\begin{equation}
E_x^r\left( {x,y} \right) = {A_x}\frac{{\partial E_x^i\left( {x,y} \right)}}{{\partial x}} + {A_y}{e^{i\Delta \phi }}\frac{{\partial E_y^i\left( {x,y} \right)}}{{\partial y}}
\label{eq:3}
\end{equation}
Furthermore, when a constructive interference between two derivatives with the same scaling factor is realized, i.e. $\left| {{A_x}} \right| = \left| {{A_y}} \right|$ and $\Delta \phi  = \arg \left( {{A_x}} \right) - \arg \left( {{A_y}} \right)$, we enable a vector-to-scalar field transformation which corresponds to the divergence computation of the incident vector fields as
\begin{equation}
E_x^r\left( {x,y} \right) = {A_x}{\nabla _ \bot } \cdot {{\bf{E}}_{{\bf{i}}}}
\label{eq:4}
\end{equation}
where ${\nabla _ \bot } = {{\bf{e}}_{\bf{x}}}\frac{\partial }{{\partial x}} + {{\bf{e}}_{\bf{y}}}\frac{\partial }{{\partial y}}$ is the transverse divergence operator.

\section{Optical divergence computation on vector fields}
To demonstrate the divergence computation, we numerically simulate such a vector-to-scalar field transformation by the finite element method using the commercial software COMSOL. Here the surface conductivity of monolayer graphene is described by the Drude model $\sigma  = \frac{{{e^2}}}{{\pi {\hbar ^2}}}\frac{{i{E_{\rm{F}}}}}{{\omega  + i{\tau ^{ - 1}}}}$, where $e$ is the elementary charge, $\tau $ is the carrier relaxation time which satisfies $\tau  = \mu {E_{\rm{F}}}/ev_{\rm{F}}^2$. ${E_{\rm{F}}}$, $\mu $ and ${v_{\rm{F}}}$ correspond to the Fermi energy, the carrier mobility and the Fermi velocity, respectively \cite{GarciadeAbajo2014}. The Fermi energy and the carrier mobility of the monolayer graphene are assumed to be ${E_{\rm{F}}}{\rm{ = }}0.6{\rm{eV}}$ and $\mu  = 1.3793 \times {10^4}{\rm{c}}{{\rm{m}}^{\rm{2}}}{{\rm{V}}^{{\rm{ - 1}}}}{{\rm{s}}^{{\rm{ - 1}}}}$ \cite{horng2011drude,bolotin2008ultrahigh}, and correspondingly the relaxation time is $\tau  = 0.8276{\rm{ps}}$. The refractive index of the silicon is ${n_{{\rm{Si}}}} = 3.4164$. The corresponding dispersion diagram of the SPP on the air-graphene-Si interface is shown in Fig.~\ref{fig:1}(b).

Here we consider that the incident wave is at 5.368THz and the corresponding wavelength is $\lambda  = 55.9{\rm{\upmu m}}$. The incident angle of paraxial beam is $\theta  = {44.427^ \circ }$. The thickness of grating is $d = 5{\rm{\upmu m}}$ and the period of grating is $\Lambda  = 4.101{\rm{\upmu m}}$ for the phase matching condition. We numerically calculate the spatial spectral transfer functions ${H_{xx}}$ and ${H_{xy}}$ for the metasurface.  The requirement of $\left| {{A_x}} \right| = \left| {{A_y}} \right|$ can be achieved by appropriately designing the geometric parameters of the gratings, with the width of grating slot $W = 1.1323{\rm{\upmu m}}$ and the height of grating slot $H = 0.8194{\rm{\upmu m}}$. 
 
\begin{figure}
\centerline{\includegraphics[width=3.2in]{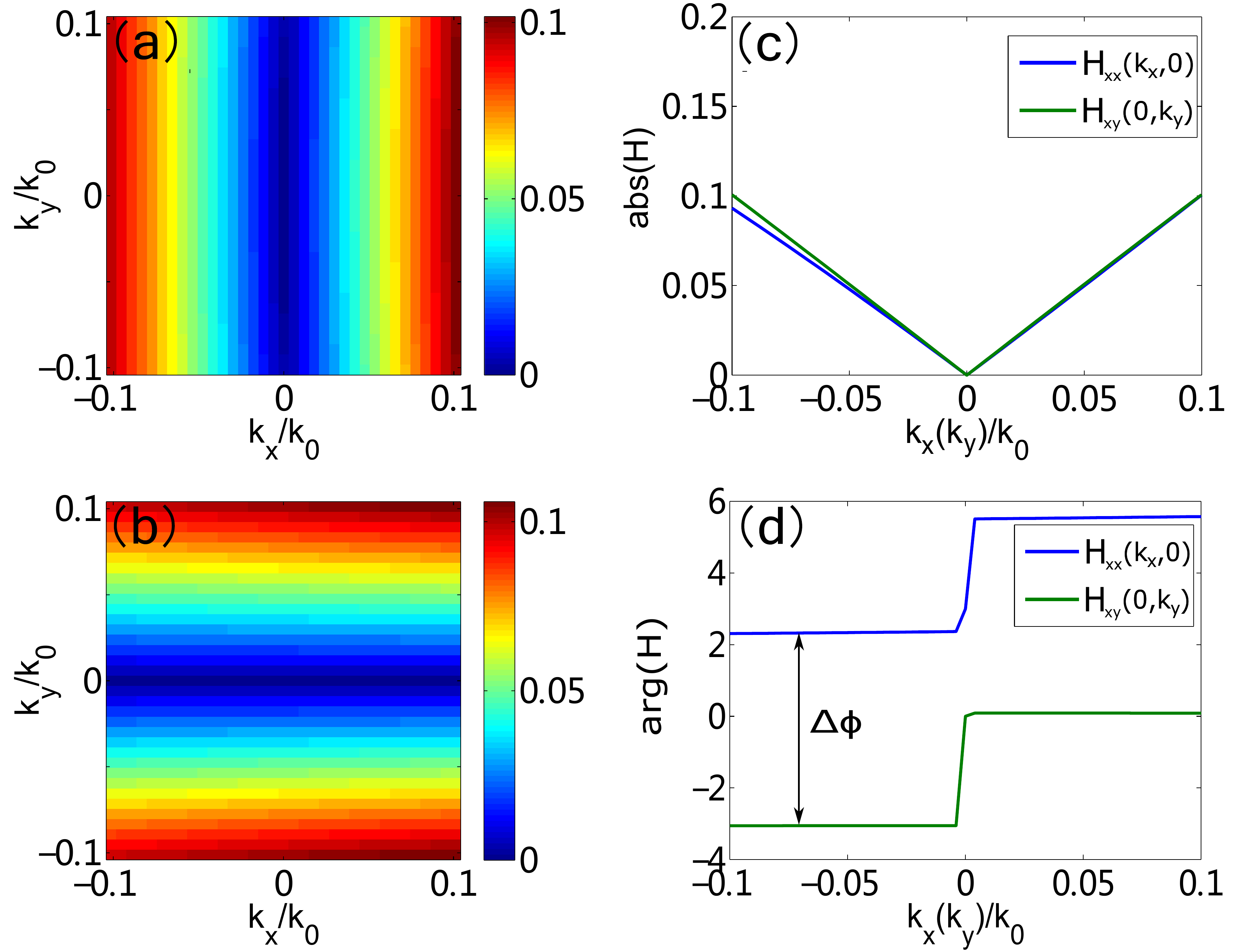}}
\caption{\label{fig:2} Numerical simulation results for the spatial spectral transfer functions of the metasurface (a) $\left| {{H_{xx}}} \right|$ and (b) $\left| {{H_{xy}}} \right|$. (c-d) Amplitude and phase distributions of ${H_{xy}}$ along ${k_x} = 0$ (green line) and of ${H_{xx}}$ along ${k_y} = 0$ (blue line).}
\end{figure}

Fig.~\ref{fig:2}(a) and (b) show that the distributions of  $\left| {{H_{xx}}} \right|$ and $\left| {{H_{xy}}} \right|$ are mainly dependent on ${k_x}$ and ${k_y}$, respectively. Furthermore, the blue (green) line in Fig.~\ref{fig:2}(c) and (d) corresponds to the amplitude and phase of ${H_{xx}}$ (${H_{xy}}$) along ${k_y} = 0$ (${k_x} = 0$). As expected by Eq.~(\ref{eq:1}) and Eq.~(\ref{eq:2}), Fig.~\ref{fig:2}(c) shows that $\left| {{H_{xx}}} \right|$ and $\left| {{H_{xy}}} \right|$ exhibit good linear dependency on ${k_x}$ and ${k_y}$, and the minimums occur at ${k_x} = 0$ and ${k_y} = 0$, respectively. More importantly, they almost overlay with each other, which indicates that the requirement of $\left| {{A_x}} \right| = \left| {{A_y}} \right|$ is approximately achieved. Fig.~\ref{fig:2}(d) shows the $\pi $ phase shift at the point ${k_{x\left( y \right)}} = 0$ for ${H_{xx}}$ and ${H_{xy}}$, which is required by the first-order differentiation operation. In addition, the invariant phase difference between ${H_{xx}}$ and ${H_{xy}}$ is  $\Delta \phi  = \arg \left( {{A_x}} \right) - \arg \left( {{A_y}} \right) = 5.4$, which can be compensated by the wave plate with appropriate thickness .

\begin{figure}
\centerline{\includegraphics[width=3.2in]{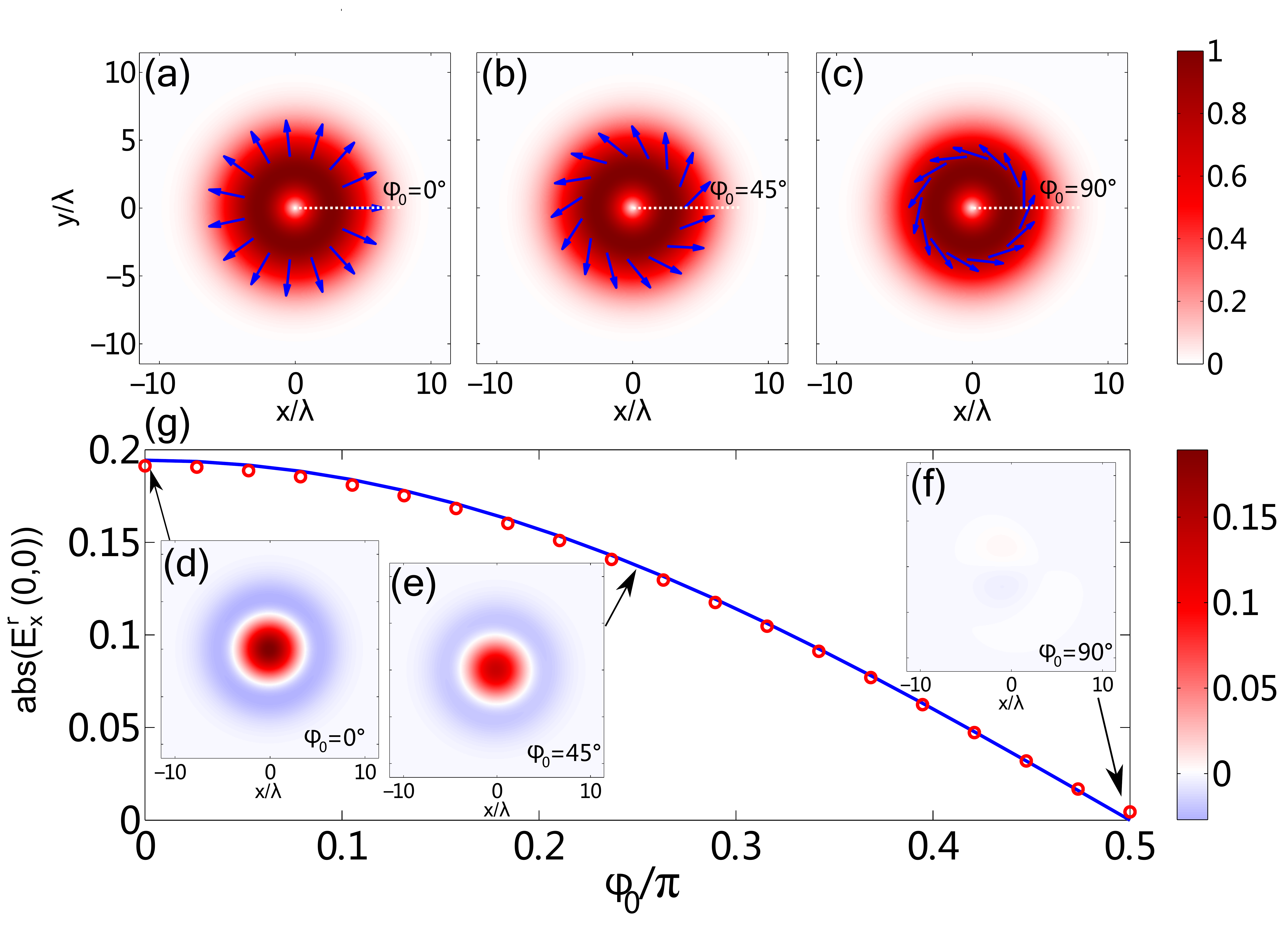}}
\caption{\label{fig:3} (a-c) Three vectorial paraxial beams with the same topological charge n=1 but the different spatial distribution of polarization states, which are determined by a initialized phase ${\varphi _0} = {0^ \circ },{45^ \circ },{90^ \circ }$, respectively. (d-f) Simulation results of the output beams correspond to the incident vector fields shown in (a-c), respectively. (g) Amplitude of the output beams at the center as the varied initialized phase ${\varphi _0}$, where the red circle and the blue solid lines correspond to the numerical simulation result and the ideal one, respectively. }
\end{figure}

To illustrate the divergence computation, we consider a series of cylindrical vector (CV) fields, which have the same intensity distribution but different distributions of polarization states as shown in Fig.~\ref{fig:3}(a-c). Such cylindrical vector fields can be generated by an interference method \cite{jones2016poincare}, with two Laguerre-Gaussian beams with opposite topological charges and opposite handedness of circularly polarization:
\begin{equation}
{{\bf{E}}_{{\bf{i}}}} = \frac{1}{{\sqrt 2 }}\left( {{\rm{LG}}_0^{ - n}{e^{ - i{\varphi _0}}}{{\bf{e}}_{\bf{ + }}} + {\rm{LG}}_0^n{e^{i{\varphi _0}}}{{\bf{e}}_ - }} \right)
\label{eq:5}
\end{equation}
Here ${\rm{LG}}_0^n$ corresponds to the Laguerre-Gaussian beam with the electric field distribution ${\rm{LG}}_0^n = {\left( {\sqrt {\frac{{2e}}{{\left| n \right|}}} \frac{r}{{{w_0}}}} \right)^{\left| n \right|}}{e^{in\varphi }}{e^{ - {{{r^2}} \mathord{\left/{\vphantom {{{r^2}} {w_0^2}}} \right.\kern-\nulldelimiterspace} {w_0^2}}}}$, $r$ and $\varphi $ are the polar coordinates of the cross section of beam. The ${\varphi _0}$ is an initialized phase of the Laguerre-Gaussian beam. The unit vectors ${{\bf{e}}_{\bf{ + }}}$ and ${{\bf{e}}_{\bf{ - }}}$ are the left- and right-hand circularly polarized states of the two Laguerre-Gaussian beams respectively. When the topological charge is fixed, the cylindrical vector fields of Eq. (\ref{eq:5}) have the same amplitude distribution $\left| {{{\bf{E}}_{{\bf{i}}}}} \right| = {\left( {\sqrt {\frac{{2e}}{{\left| n \right|}}} \frac{r}{{{w_0}}}} \right)^{\left| n \right|}}{e^{ - {{{r^2}} {\left/{\vphantom {{{r^2}} {w_0^2}}} \right.} {w_0^2}}}}$. Here we first choose the topological charge in Laguerre-Gaussian beam as $n = 1$, and the radius of the beam waist is ${w_0} = 3.82{\lambda}$. The amplitude distribution of incident field $\left| {{{\bf{E}}_{{\bf{i}}}}} \right|$ is shown in Fig.~\ref{fig:3}(a-c). On the other hand, the polarization state of the CV fields is only dependent on the azimuthal angle $\varphi$:
\begin{equation}
{\bf{e}}\left( \varphi  \right) = \cos \left[ {\left( {n - 1} \right)\varphi  + {\varphi _0}} \right]{{\bf{e}}_{\bf{r}}} + \sin \left[ {\left( {n - 1} \right)\varphi  + {\varphi _0}} \right]{{\bf{e}}_{{\upvarphi }}}
\label{eq:6}
\end{equation}
where ${{\bf{e}}_{\bf{r}}}$ and ${{\bf{e}}_{{\upvarphi }}}$ are the orthogonal bases associated with the polar coordinates $\left( {r,\varphi } \right)$. Therefore, in the case of $n = 1$, the polarization state ${\bf{e}}$ has a uniform angle ${\varphi _0}$ from the basis ${{\bf{e}}_{\bf{r}}}$, shown as the arrows in Fig.~\ref{fig:3}(a-c), with different initialized phase factors ${\varphi _0} = {0^ \circ }$, ${45^ \circ }$, and ${90^ \circ }$, respectively. We note that at the center of the beam, the orientation of resultant vector field is undefined and the amplitude of the electric field vanishes, which corresponds to a vector point singularity (V-point) \cite{freund2002polarization,burresi2009observation,soskin2003optical}.

We numerically simulate the output beams after analyzing the $x$ polarization component of reflected light by the metasurface. Fig.~\ref{fig:3}(d-f) correspond to the simulation results of the output field for the incident CV fields in Fig.~\ref{fig:3}(a-c). According to Eq. (\ref{eq:5}), the ideal result of the divergence computation is $\frac{{2\sqrt {2e} {A_x}}}{{{w_0}}}\cos {\varphi _0}\left( {1 - {{\left( {{r {\left/{\vphantom {r {{w_0}}}} \right.} {{w_0}}}} \right)}^2}} \right){e^{ - {{{r^2}}{\left/{\vphantom {{{r^2}} {w_0^2}}} \right.} {w_0^2}}}}$, which has a zero value at $r = {w_0}$, and agrees well with the simulation results shown in Fig.~\ref{fig:3}(d-f). Although the intensities of the incident vector fields are the same, the divergence computation results exhibit different amplitudes, dependent on the polarization distribution of the incident field. Fig.~\ref{fig:3}(g) shows the numerical simulation result (red circles) and the ideal one of the divergence computation (blue solid line) around the beam centers, which agree well with high accuracy. Especially in this case of ${\varphi _0} = {\rm{90}}^\circ $ [Fig.~\ref{fig:3}(c)], the divergence is absent since the incident vector fields are purely rotational.

\begin{figure}
\centerline{\includegraphics[width=3.2in]{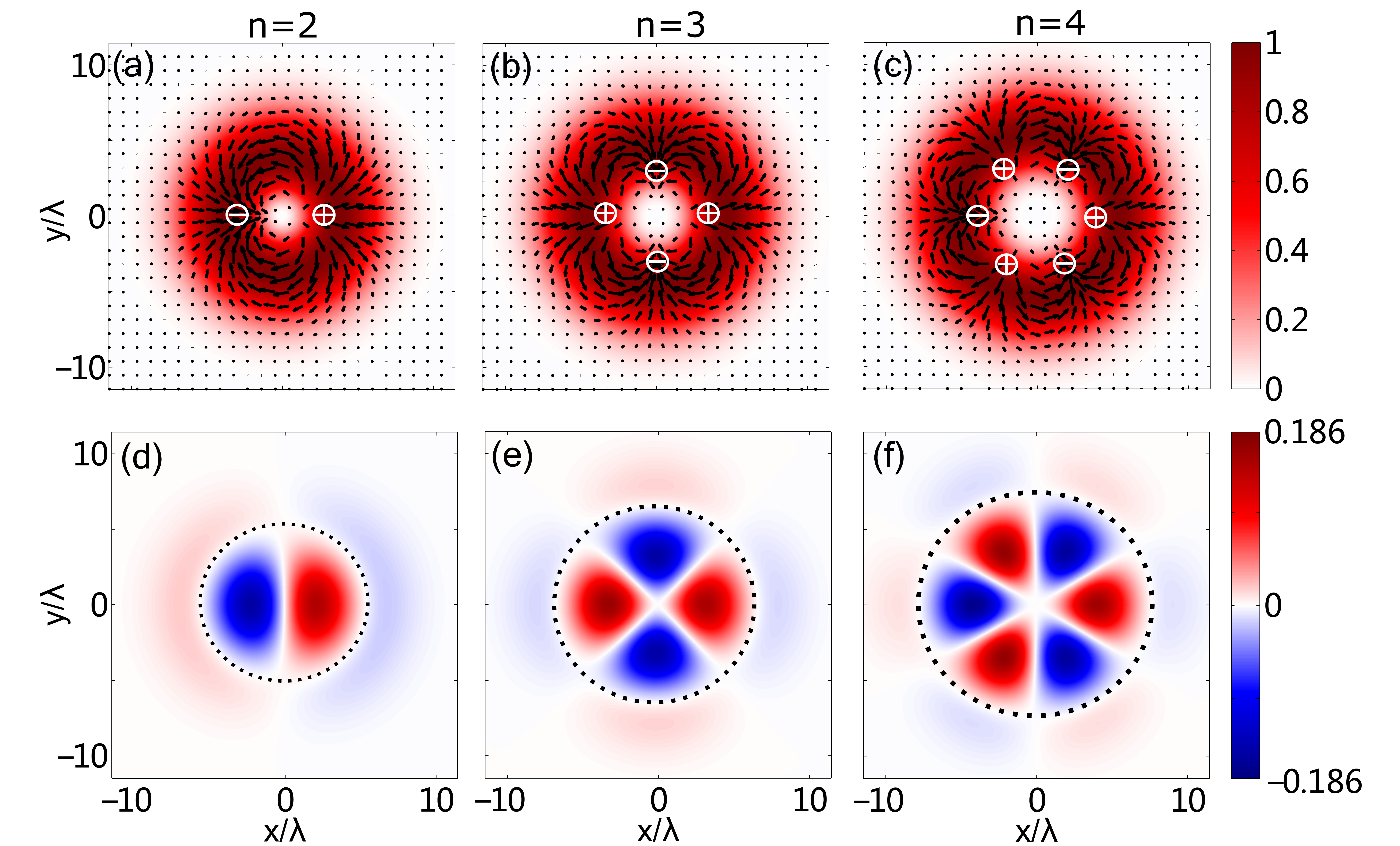}}
\caption{\label{fig:4} Divergence operation for cylindrical vector  beam with different topological charges. (a-c) incident electric vector field distribution with topological charge $n = 2,3,4$ respectively. The sign $ \pm $ represents the positive and negative equivalent charge density. (d-f)  Simulation results of the output beams correspond to the incident vector fields shown in (a-c), respectively. The black dashed circle line corresponds to the position where reflected electric field vanishes.}
\end{figure}

Conventionally, the concepts of source and sink are used to describe the polarization distributions in vector field, respectively, with positive and negative discrete charges. By computing the divergence, the discrete charges should redistribute themselves in a continuous manner. Indeed, the intensity image obtained by our device, shown in Fig. ~\ref{fig:3}(d-f), present the density distribution of the equivalent charges in vectorial paraxial beams.

The density of equivalent charge distribution exhibits more complicated forms if topological charge of incident Laguerre-Gaussian beam has higher orders. Fig.~\ref{fig:4}(a)-(c) correspond to the cylindrical vector fields of Eq. (\ref{eq:5}), with different topological charges $n = 2$, $3$, and $4$. The arrows in Fig.~\ref{fig:4}(a-c) exhibit the distributions of the polarization states. The sources and sinks of the polarization distributions are represented by the sign $ \pm $ in Fig.~\ref{fig:4}(a-c). Fig.~\ref{fig:4}(d-f) are the simulation results of the optical computation of the divergence, performed by the proposed metasurface. In a good agreement with the polarization states as shown in Fig.~\ref{fig:4}(a-c), the amplitude of the output field quantitatively models the density distribution of positive and negative charges. More interestingly, Fig.~\ref{fig:4}(d-f) intuitively show that out of central area (the dashed circles) there are opposite charge distributions, which is not so obvious to infer from the polarization states in Fig.~\ref{fig:4}(a-c) directly.

\section{Conclusion}

We propose a periodic plasmonic metasurface to realize the optical computation of divergence operation. Moreover, we demonstrate the divergence operations for a series of cylindrical vector fields, which extract the equivalent charge distributions with different polarization states or topological charges. Our study shows that such optical computing devices provide a new direct pathway to elucidate specific polarization singularities in vector fields. It would be more helpful in the three-dimensional cases, which require much more stable measurement \cite{flossmann2008polarization,flossmann2005polarization}. Also such real-time high-throughput analyzation of vector field provides possible applications in microscopic imaging, target recognition, and augmented reality.

The authors acknowledge funding through National Natural Science Foundation of China (NSFC Grants No. 91850108 and No. 61675179), National Key Research and Development Program of China (Grant No.2017YFA0205700), the Open Foundation of the State Key Laboratory of Modern Optical Instrumentation, and the Open Research program of Key Laboratory of 3D Micro/Nano Fabrication and Characterization of Zhejiang Province.


%

\end{document}